\documentclass[a4paper,11pt]{article}
\usepackage{longtable}
\pdfoutput=1 

\usepackage{jcappub} 

\usepackage[T1]{fontenc} 

\title{\boldmath The impact of FRB dispersion measure probability distribution functions on cosmographic estimates}

\author[a,1]{Thais Lemos,\note{Corresponding author}}
\author[b,a]{Rodrigo Gon\c{c}alves,}
\author[a]{Jailson Alcaniz}

\affiliation[a]{Observat\'orio Nacional, Rio de Janeiro - RJ, 20921-400, Brasil}
\affiliation[b]{ Departamento de F\'{\i}sica, Universidade Federal Rural do Rio de Janeiro, Serop\'edica - RJ, 23897-000, Brasil}

\emailAdd{thaislemos@on.br}
\emailAdd{rsg\_goncalves@ufrrj.br}
\emailAdd{alcaniz@on.br}

\abstract{Recent cosmological observations have reopened the discussion about the model that best describes the dynamics of the Universe, highlighting the need for cosmological model-independent analyses. In this paper, we utilize the cosmographic approach applied to a robust sample of 106 well-localized Fast Radio Bursts (FRBs) within the redshift range $z \le 0.7$ to constrain the Hubble constant $H_0$, the deceleration parameter $q_0$, and the jerk parameter $j_0$. Our primary goal is to assess the impact of intergalactic medium (IGM) inhomogeneities on cosmographic parameter estimation. To this end, we consider the statistical behavior of these parameters under two distinct functional forms for the IGM dispersion measure ($\mathrm{DM_{IGM}}$) probability density function (PDF): a Gaussian distribution (Distribution I) and a quasi-Gaussian distribution (Distribution II) that accounts for the skewed structure of cosmic large-scale environments along the lines of sight. We further investigate the role of the baryon mass fraction by considering both fixed and free-parameter scenarios. We find that the inferred cosmographic constraints, particularly those on $q_0$, depend sensitively on both the assumed IGM distribution and the adopted parameter priors.

}

\begin{document}
\maketitle
\flushbottom


\section{Introduction}
\label{sec:intro}


Recent Baryon Acoustic Oscillations (BAO) measurements from the Dark Energy Spectroscopic Instrument (DESI) \cite{DESI} have intensified the debate on the nature of dark energy by suggesting a preference for an evolving dark energy component. These observations have yielded increasingly stringent constraints on the dark energy equation of state (EoS) across different $w(a)$ parameterizations~\cite{Chevallier:2000qy,Linder:2002et,Barboza:2008rh}. In addition, the discrepancy in the Hubble constant ($H_{0}$) measurements between early-time observations from the Cosmic Microwave Background, $H_{0} = 67.36 \pm 0.54$ km s$^{-1}$ Mpc$^{-1}$ reported by the Planck collaboration \cite{Planck2018}, and late-time measurements from Cepheid-calibrated Type Ia supernovae (SNe), $H_{0} = 73.04 \pm 1.01$ km s$^{-1}$ Mpc$^{-1}$ from the SH0ES collaboration \cite{Riess2022}, further highlights tensions within the $\Lambda$CDM model. In this context, these discrepancies motivate the use of model-independent approaches to constrain cosmological parameters.

One way to probe the late-time evolution of the Universe without assuming a specific cosmological model is through cosmography \cite{Weinberg:1972kfs,Chiba1998,Visser2004,Lazkoz:2013ija,Rodrigues:2025tfg}. This approach provides a model-independent description of the cosmic expansion history, relying solely on the space-time geometry defined by the Friedmann–Lemaître–Robertson–Walker (FLRW) metric, without invoking the underlying dynamical equations. Using Taylor expansion, one can estimate the Hubble constant, deceleration, jerk, and other higher-order parameters directly from observations. 


Fast Radio Bursts (FRBs) constitute a promising observational probe for such studies. These highly energetic radio transients, characterized by millisecond durations and frequencies of order GHz (for reviews, see \cite{Thornton2013,Petroff2015,Petroff2016,Petroff2022}), exhibit large dispersion measures ($\mathrm{DM}$) that strongly suggest an extragalactic, and often cosmological, origin \cite{Platts2019}. Although their progenitors and emission mechanisms remain uncertain (see \cite{Dolag2015}), the growing sample of well-localized FRBs has opened new avenues for precision cosmology. The first event was observed in 2007 by the Parkes telescope \cite{Lorimer2007}, and since then several events have been detected by new surveys, especially by the Canadian Hydrogen Intensity Mapping Experiment (CHIME, \cite{CHIME}), which recently reported its second catalog with $> 4000$ observed events \cite{CHIME2}. However, only a few of these events in the literature are well localized, i.e. with the correspondent redshift. When the redshift of the host galaxy is determined, one can combine the $\mathrm{DM}$ with the redshift and use it as an astrophysical and cosmological probe (see \cite{Walters2018,Wei2018,Lin2021,Wu2021,Lemos2023,Lemos2025} for applications of FRBs in cosmology).

In practice, several issues limit the cosmological applications of FRBs. The first is related to the inhomogeneous electron distribution in the intergalactic medium, which introduces variance in the dispersion measure \cite{Macquart2020,Takahashi2021}, often of the same order of magnitude as the dispersion measure itself. These density fluctuations cannot be directly inferred from observations or accurately modeled because of the difficulty of probing the electron distribution along the FRB pulse propagation path. For this reason, these fluctuations can be treated either as a probability distribution \cite{Macquart2020} or as a fixed contribution in the statistical analysis \cite{Takahashi2021}. Another limitation arises from the host galaxy contribution, which is difficult to observe and characterize due to its complex and poorly constrained nature \cite{Xu2015}. To mitigate this issue, it is common to assume a log-normal distribution~\cite{Walker2020}. Finally, an additional limitation arises from the poor observational knowledge of the evolution of the baryon fraction in the intergalactic medium ($f_{\mathrm{IGM}}$) \cite{Shull2012,Meiksin2009}, which is degenerate with cosmological parameters.

In this work, we investigate the impact of different probability density functions for the FRB dispersion measure (DM) on the constraints of the cosmographic parameters, namely the Hubble constant,deceleration and jerk parameters using a sample of 106 well-localized FRBs. We consider two functional forms for the dispersion measure PDF of FRBs: the first assumes a Gaussian distribution, in which the fluctuations represent the variance in DM produced by inhomogeneities in the electron density of the intergalactic medium; the second adopts a quasi-Gaussian distribution characterized by a long right-hand tail, which accounts for the asymmetric distribution of electron density fluctuations in the IGM. 

This paper is organized as follows. In Section~\ref{sec:frb_theory}, we review the theoretical background of FRB dispersion measures. In Section~\ref{sec:cosmohraphy}, we introduce the cosmographic framework and derive the dispersion measure components within this approach. The data and methodology are described in Section~\ref{sec:methodology}. In Sections~\ref{sec:results} and \ref{sec:newprior}, we present the results of our analysis. Finally, in Section~\ref{sec:conclusion}, we summarize our main conclusions.

\section{FRB's theory}
\label{sec:frb_theory}

The observed dispersion measure ($\mathrm{DM_{obs}}$) of a FRB quantifies the frequency-dependent time delay imparted by free electrons along the line of sight. It can be decomposed into contributions from several distinct media \cite{Deng2014, Gao2014}:

\begin{equation}\label{eq:dm}
\mathrm{DM}_{\mathrm{obs}} = \sum_{i} \mathrm{DM}_{i} (z),
\end{equation}
where the index $i$ corresponds to the Milky Way's interstellar medium (ISM), the Milky Way halo (halo), the intergalactic medium (IGM), and the host galaxy of the FRB (host).

The total observed $\mathrm{DM}_{\mathrm{obs}}(z)$ is a direct measurement from the FRB event. The Galactic interstellar contribution, $\mathrm{DM}_{\mathrm{ISM}}$, can be estimated using established models of the Galactic electron distribution derived from pulsar observations \cite{Taylor1993, Cordes2002, Yao2017}. For the contribution from the Milky Way's halo, which is less constrained, we adopt a fixed value of $\mathrm{DM}_{\mathrm{halo}} = 50$ pc/cm$^{3}$ \cite{Macquart2020}.

The host galaxy's contribution remains particularly challenging to model. Its value depends on factors such as host galaxy type, the orientation of the FRB source within it, and the properties of the local plasma \cite{Xu2015}. Therefore, one can model it as:

\begin{equation}
\mathrm{DM}_{\mathrm{host}} (z) = \frac{\mathrm{DM}_{\mathrm{host,0}}}{1+z},
\end{equation}
where the term $(1+z)$ accounts for cosmic dilation \cite{Deng2014, Ioka2003}, and $\mathrm{DM}_{\mathrm{host,0}}$ is the host galaxy's DM contribution in its rest frame. 

The final and dominant contribution is from the intergalactic medium (IGM). The average IGM dispersion measure is given by \cite{Deng2014}:

\begin{equation}\label{eq:igm}
\mathrm{DM}_{\mathrm{IGM}}(z) = \frac{3c\Omega_{b}H_{0}^{2}}{8\pi Gm_{p}} \int_{0}^{z} \frac{(1+z')f_{\mathrm{IGM}}(z')\chi(z')}{H(z')} dz' \; ,
\end{equation}
where $c$ is the speed of light, $\Omega_{b}$ is the present-day baryon density parameter, $H_{0}$ is the Hubble constant, $G$ is the gravitational constant, $m_{p}$ is the proton mass, $f_{\mathrm{IGM}}(z)$ is the baryon fraction in the IGM, and $H(z)$ is the Hubble parameter. The term $\chi(z) = Y_{H}\chi_{e,H}(z) + Y_{He}\chi_{e,He}(z)$ represents the free electron number fraction per baryon, with $Y_{H} = 3/4$ and $Y_{He} = 1/4$ being the mass fractions of hydrogen and helium, respectively. For redshifts $z < 3$, hydrogen and helium are fully ionized \cite{Meiksin2009, Becker2011}, giving $\chi_{e,H}(z) = \chi_{e,He}(z) = 1$.

From Eq. \ref{eq:dm}, the extragalactic dispersion measure used in the statistical analysis is defined as
\begin{equation}
\label{eq:ext_obs}
\mathrm{DM}^{\mathrm{obs}}_{\mathrm{ext}}(z)=\mathrm{DM}_{\mathrm{obs}}(z)-\mathrm{DM}_{\mathrm{MW}},
\end{equation}
which represents the observed contribution. The corresponding theoretical extragalactic dispersion measure is
\begin{equation}
\label{eq:ext_th}
\mathrm{DM}^{\mathrm{th}}_{\mathrm{ext}}(z)=\mathrm{DM}_{\mathrm{host}}(z)+\mathrm{DM}_{\mathrm{IGM}}(z).
\end{equation}


\section{Cosmographic approach}
\label{sec:cosmohraphy}

The cosmography is a model-independent approach to constrain the kinematics equations of the Universe directly from observations, and is independent of any assumption about the underlying dynamics. This approach requires the FLRW metric, which is described as

\begin{equation}
    dS^{2} = -c^{2} dt^{2} + a^{2}(t) \left[ \frac{dr^{2}}{1-kr^{2}} + r^{2} (d\theta ^{2} + \mathrm{sin}^{2}\theta \,d\phi ^{2}) \right],
\end{equation}
where $k = -1, 0$ and $+1$ are the curvature of space for open, flat, and closed geometries of the Universe, respectively, and here we consider $k = 0$.

In order to apply the cosmographic approach in the DM of FRBs, we can define the cosmographic functions in terms of scale factor ($a$) \cite{Chiba1998,Visser2004}

\begin{align}
    H (t)  &=  \frac{1}{a} \frac{da}{dt},\label{eq:H}\\
    q (t)  &= - \frac{1}{a} \frac{d^{2}a}{dt^{2}} \left(\frac{1}{a} \frac{da}{dt} \right)^{-2},\label{eq:q}\\
    j (t) &= \frac{1}{a} \frac{d^{3}a}{dt^{3}} \left(\frac{1}{a} \frac{da}{dt} \right)^{-3},\label{eq:j}\\
    s (t) &= \frac{1}{a} \frac{d^{4}a}{dt^{4}} \left(\frac{1}{a} \frac{da}{dt} \right)^{-4},\label{eq:s}
\end{align}
where $H(t)$, $q(t)$, $j(t)$, and $s(t)$ denote the Hubble, deceleration, jerk, and snap parameters, respectively. The Hubble parameter $H(t)$ characterizes the cosmic expansion rate, while the deceleration parameter $q(t)$ indicates whether the expansion is accelerating ($q<0$) or decelerating ($q>0$). The jerk parameter $j(t)$ is sensitive to changes in the acceleration of the expansion and can be used to probe possible transitions in the cosmic expansion history. Finally, the snap parameter $s(t)$ provides additional discriminatory power between cosmological models with a cosmological constant and those that allow for an evolving dark energy component.

Performing a Taylor expansion of the Hubble parameter (Eq. \ref{eq:H}) around the present time ($t_{0}$) and using the relation between cosmic time and redshift ($dt = -[(1+z)H(z)]^{-1}dz$), we obtain 

\begin{equation}\label{eq:Hz_series}
 H(z) = H_{0} \left\{ 1 + (1+q_{0})z + \left(j_{0} - q_{0}^{2}\right)\frac{z^{2}}{2} + \left[3q_{0}^{2} (1+q_{0})-j_{0}(3+4q_{0}) - s_{0} \right]\frac{z^{3}}{6}  + \mathcal{O}(z^{4}) \right\},    
\end{equation}
being $q_{0}$, $j_{0}$ and $s_{0}$ the parameters from Eqs. \ref{eq:q}, \ref{eq:j} and \ref{eq:s} at present time. 

Using the equation above and considering $f_{\mathrm{IGM}} (z)$ constant on redshift ($f_{\mathrm{IGM}}(z) = f_{\mathrm{IGM,0}}$), we can Taylor expand the average of $\mathrm{DM_{IGM}} (z)$ (Eq. \ref{eq:igm}) as

\begin{align}\label{eq:igmz_series}
 \mathrm{DM_{IGM}} (z) &= A_{\mathrm{IGM}} \,f_{\mathrm{IGM,0}} \, \Omega_{b} \,H_{0} \bigg\{z - q_{0}\frac{z^{2}}{2} + \left[ q_{0} (2+3q_{0}) -j_{0}\right]\frac{z^{3}}{6} + \big[ -15q_{0}^{3} - 18q_{0}^{2} -6q_{0} \nonumber\\
&+2j_{0}(3+5q_{0}) + s_{0} \big]\frac{z^{4}}{24} + \mathcal{O}(z^{5}) \bigg\} \, ,
\end{align}
being $A_{\mathrm{IGM}}=21c / (64\pi G m_{p})$. Note that this equation agrees with the equations reported by \cite{Gao2024} (when $\alpha = 0$) and \cite{Lazaro}. However, this equation disagrees at the second-order from Ref. \cite{Fortunato2023}.

The Taylor expansions in Eqs.~\ref{eq:Hz_series} and \ref{eq:igmz_series} do not converge at high redshifts ($z>1$). This issue can be addressed by reparameterizing the redshift as $ y = z/(1+z)$. In terms of the $y$-redshift, the resulting Taylor series are well behaved, ensuring the convergence of both $H(y)$ and $\mathrm{DM}_{\mathrm{IGM}}(y)$ at high redshift. In terms of $y$-redshift, the Hubble parameter can be expressed as

\begin{align}\label{eq:Hy_series}
     H(y) &= H_{0} \bigg\{ 1 + (1+q_{0})y + \left(2+2q_{0} + j_{0} - q_{0}^{2}\right)\frac{y^{2}}{2} + \big[3q_{0}^{2}(q_{0}-1) + 2q_{0}(3-2j_{0})+3j_{0}    \nonumber\\
     &-s_{0}+ 6\big]\frac{y^{3}}{6} + \big[ 36q_{0}^{3} + 24q_{0} (1-2j_{0})-12s_{0}+24\big] \frac{y^{4}}{24}+ \mathcal{O}(y^{5}) \bigg\}.
\end{align}

Therefore, the IGM contribution in terms of the $y$-series can be written as 

\begin{align}\label{eq:igmy_series}
 \mathrm{DM_{IGM}} (y) &= A_{\mathrm{IGM}} \,f_{\mathrm{IGM,0}} \, \Omega_{b} \,H_{0} \bigg\{y +(2- q_{0})\frac{y^{2}}{2} + \left[ 3q_{0}^{2} - 4q_{0} - j_{0} + 6\right]\frac{y^{3}}{6} + \big[ -15q_{0}^{3}  \nonumber\\
 & + 18q_{0}^{2} - 18q_{0} + 2j_{0}(5q_{0}-3)+s_{0} + 24\big]\frac{y^{4}}{24} + \mathcal{O}(z^{5}) \bigg\}.   
\end{align}

\begin{figure*}
\begin{center}
\includegraphics[width=0.7\textwidth]{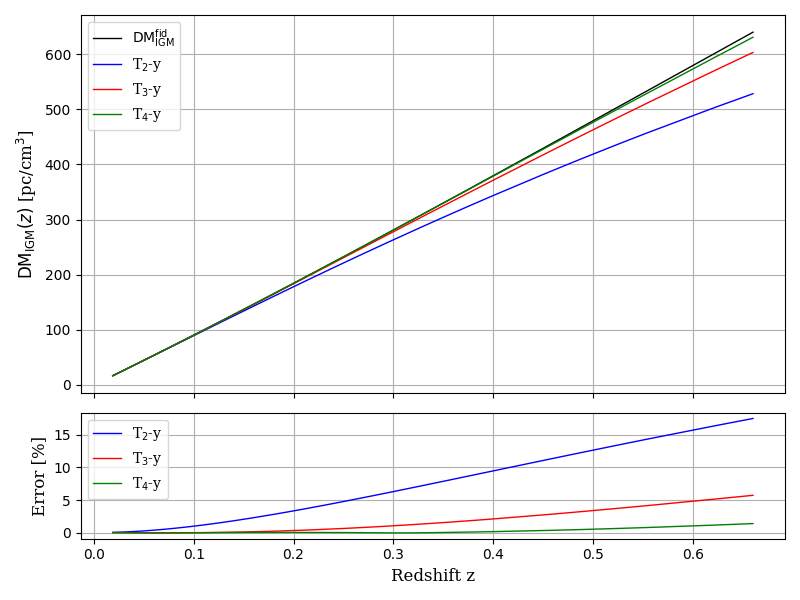}
\caption{Intergalactic dispersion measure $\mathrm{DM_{IGM}}$ as a function of redshift (top panel) and the relative errors between the cosmographic approximation and the fiducial model (bottom panel). The black line represents the $\mathrm{DM_{IGM}}$ calculated from Eq. \ref{eq:igm}, whereas the blue, red, and green curves correspond to the cosmographic expansions in the $y$-redshift, truncated at second, third, and fourth order, respectively (Eq. \ref{eq:igmy_series}). }
\label{fig:IGM}
\end{center} 
\end{figure*}

In Fig.~\ref{fig:IGM}, we illustrate the performance of the cosmographic approach in reproducing the intergalactic dispersion measure, $\mathrm{DM_{IGM}}$. We compare the second-, third-, and fourth-order cosmographic expansions with the fiducial $\mathrm{DM_{IGM}}$ model given by Eq.~\ref{eq:igm}, assuming a flat $\Lambda$CDM cosmology with parameters $H_{0}=67.35\pm0.54$, $\Omega_{m}=0.3153\pm0.0073$, and $\Omega_{b}h^{2}=0.02237\pm0.00015$, as reported by the Planck Collaboration \cite{Planck2018}. We further adopt $f_{\mathrm{IGM,0}}=0.83$ and $\mathrm{DM_{host,0}}=120$ pc/cm$^{-3}$.

The second-order expansion starts to deviate significantly from the fiducial $\Lambda$CDM prediction at $z\gtrsim0.2$, reaching relative errors above $15\%$ for $z\gtrsim0.6$. In contrast, the third- and fourth-order expansions provide a much better agreement, with relative errors $\lesssim5\%$ over the same redshift range. We choose to truncate the cosmographic expansion at third order in the remainder of this work, given that the improvement from third to fourth order is modest (of order $\sim4\%$ in relative error), while the latter introduces an additional free parameter and increased degeneracies.

\section{Data and methodology}
\label{sec:methodology}

\subsection{Data}

The currently available sample of well-localized FRBs comprises 117 events (for details of the FRB catalog\footnote{https://blinkverse.alkaidos.cn}, see \cite{Blinkverse}). However, several events are excluded from our analysis for the following reasons: FRB 20240123A \cite{FRB220204A}, FRB 20221029A \cite{FRB220204A}, FRB 20220610A \cite{FRB20220610A}, and FRB 20230521B \cite{FRB220204A}, which lie at redshifts $z=0.968$, $0.975$, $1.016$, and $1.354$, respectively, while our analysis is restricted to $z \leq 0.7$; FRB 20171020A \cite{FRB171020A} and FRB 20181030 \cite{Bhardwaj2021_2}, located at $z=0.0087$ and $z=0.0039$, respectively, since cosmological effects are expected to be negligible at such low redshifts; FRB 20190520B \cite{FRB190520B}, which presents a host-galaxy contribution significantly larger than the other events; FRB 20190614 \cite{FRB190614}, which has no measurement of spectroscopic redshift and can be associated with more than one host galaxy; FRB 20200120E \cite{Bhardwaj2021}, which is likely associated with the nearby galaxy M81 at a distance of $\sim 3.6$ Mpc, although a Milky Way halo origin cannot be ruled out; and FRB 20210405I \cite{FRB210405I} and FRB 20220319D \cite{FRB220207C}, whose estimated Milky Way contributions exceed the observed dispersion measure, leading to $\mathrm{DM}_{\mathrm{ext}}^{\mathrm{obs}} < 0$.

\begin{figure*}
\begin{center}
\includegraphics[width=0.7\textwidth]{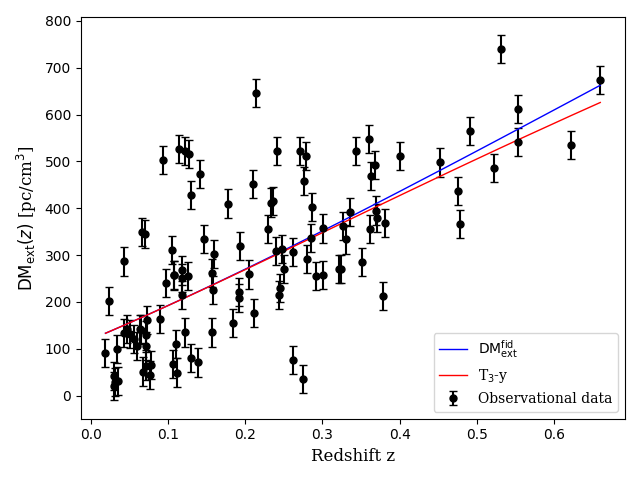}
\caption{Extragalactic dispersion measure $\mathrm{DM_{ext}}$ as a function of redshift. The black dots are the observational data, and the blue and red curves represent the fiducial $\mathrm{DM_{ext}}$ calculated from Eq. \ref{eq:ext_th} and the cosmographic expansions in the $y$-redshift, truncated at third order, respectively (Eq. \ref{eq:igmy_series}). }
\label{fig:data}
\end{center} 
\end{figure*}

The resulting working sample contains 106 FRBs, listed in Table~\ref{tab:data} \cite{FRB121102,FRB20191228,FRB180814,FRB180916,FRB180924,FRB181112,FRB181220A,FRB190102,FRB190110C,FRB190523_1,FRB190523_2,FRB190608,FRB201123A,FRB201124,FRB210117A,FRB210320,FRB210410D,FRB210603A,FRB210807D,FRB211203C,FRB220204A,FRB220207C,FRB220529A,FRB220717A,FRB220912A,FRB221219A,FRB230203A,FRB240114A,FRB240209A}. The table includes the main properties of each event: redshift, Galactic interstellar medium contribution ($\mathrm{DM}_{\mathrm{MW,ISM}}$) estimated using the NE2001 model \cite{Cordes2002}, observed dispersion measure ($\mathrm{DM}_{\mathrm{obs}}$), associated uncertainty ($\sigma_{\mathrm{obs}}$), and references. The symbol $\dagger$ in place of $\sigma_{\mathrm{obs}}$ indicates events for which no uncertainty measurement is available. For these 38 FRBs, we assign an uncertainty drawn randomly from a Gaussian distribution defined by the mean and standard deviation of the $\sigma_{\mathrm{obs}}$ distribution of the remaining events. We also displayed the working sample in Fig \ref{fig:data}, where we calculated the error bars using the observed and Galactic uncertainties.

\newpage

\subsection{Methodology}

Since the distribution of ionized plasma in the intergalactic medium is inhomogeneous, it induces variations in the IGM dispersion measure, which cannot be fully predicted by theory or directly inferred from observations. These variations can be treated either by assuming fixed fluctuation values ($\delta$) added to a Gaussian probability density function (PDF) for the IGM, or by adopting a non-Gaussian IGM PDF that introduces an asymmetry to capture the underlying fluctuations better. In the present paper, we explore both approaches as follows.


In the first approach, we consider a Gaussian distribution for the IGM, which we refer to as Distribution I, given by:

\begin{equation}\label{eq:pdf_1}
    \mathrm{P^{Dist  \; I}_{IGM}} \left( \mathrm{DM_{IGM} } \right) = \frac{1}{\sqrt{2 \pi}\sigma_{\mathrm{DM}}} \exp{\left[ - \frac{\left( \mathrm{DM_{IGM}}(z_{i},\theta) - \mathrm{DM_{IGM}} \right)^{2}}{2\sigma_{\mathrm{DM}}^{2}}\right]},
\end{equation}
with the total variance for the Dispersion Measure given by

\begin{equation}\label{eq:sigma}
    \sigma_{\mathrm{DM}}^{2} = \sigma_{\mathrm{obs}}^{2} + \sigma_{\mathrm{MW}}^{2} + \delta^{2} \;,
\end{equation}
where $\sigma_{\mathrm{obs}}$ is taken from Table \ref{tab:data},  $\sigma_{\mathrm{MW}}$ is the average galactic uncertainty assumed to be 30 pc/cm$^{3}$ \cite{Manchester2005} and $\delta$ represents $\mathrm{DM}$ fluctuations. Hydrodynamic simulations predict typical values of $\delta$ in the range $100$–$400,\mathrm{pc,cm^{-3}}$ at redshifts $z\simeq0.5$–$1$ \cite{McQuinn2014}. Following Refs. \cite{Takahashi2021,Lemos_2023a}, we adopt $\delta = 230\sqrt{z}$ pc/cm$^{3}$

For the second approach, which we refer to as Distribution II, we adopt a quasi-Gaussian PDF that allows for asymmetric fluctuations in the IGM, 

\begin{equation} \label{eq:pdf_2}
\mathrm{P^{Dist  \; II}_{IGM}}(\Delta \mid x, \mu, \sigma_{\mathrm{IGM}}) = 
A \;
\Delta^{-\beta} \;
\exp\!\Bigg[ -\frac{\mu^{2} \big( \Delta^{-\alpha} - 1 \big)^{2}}{2\alpha^{2}\sigma_{\mathrm{IGM}}^{2}} \Bigg], \; \Delta>0,
\end{equation}
which agrees with the observed distributions of $\mathrm{DM_{IGM}}$ in both semi-analytic models and hydrodynamic simulations, where $A$ is the normalization factor, $\Delta \equiv x/\mu = \mathrm{DM_{IGM}}(z_{i})/\left< \mathrm{DM_{IGM}}(z_{i}) \right> $, and $\sigma_{\mathrm{IGM}}$ is the standard deviation. For small $\sigma_{\mathrm{IGM}}$, the distribution approaches a Gaussian distribution, while for large $\sigma_{\mathrm{IGM}}$, this PDF captures the large skew that results from a few large structures that contribute significantly to the $\mathrm{DM_{IGM}}$ of many sightlines.  The parameters $\alpha$ and $\beta$ are fixed and do not change with redshift.

It is worth to note that Eq. \ref{eq:pdf_2} is very similar to the functional forms proposed by Refs. \cite{Macquart2020} and \cite{Konietzka2025}, with differences primarily in the definition of the parameters $\sigma$ and $\mu$. In our work, $\mu$ is the mean of the distribution and $\sigma$ corresponds to the standard deviation. Following Ref. \cite{Macquart2020} $\sigma$ can be estimated by $\sigma_{\mathrm{IGM}} = Fz^{-0.5}$, where $F$ quantifies the strength of the baryon feedback and can be fixed as $0.32$. In Reference \cite{Macquart2020}, the authors set $\alpha$ and $\beta$ equal to 3. We adopt the same functional form for Eq. \eqref{eq:pdf_2} and refer to it as Distribution II. 

Regarding the host-galaxy contribution, although it cannot be directly observed or individually modeled, several studies have characterized its statistical behavior using different functional forms. For instance, Ref. \cite{Walker2020} inferred a log-normal distribution with a median of $100$  pc/cm$^{3}$, while Ref. \cite{Macquart2020} found the same functional form, with the median treated as a free parameter in the range $20 - 200$ pc/cm$^{3}$. Following Refs. \cite{Walker2020,Macquart2020}, we model $\mathrm{DM}_{\mathrm{host}}$ with a log-normal distribution, given by

\begin{equation}\label{eq:pdf_3}
    \mathrm{P_{host}} \left( \mathrm{DM_{host}} | \mu, \sigma_{host} \right) = \frac{1}{\sqrt{2 \pi} \mathrm{DM_{host}} \sigma_{\mathrm{host}}} \exp{\left[ - \frac{\left( \ln{\mathrm{DM_{host}}} - \mu \right)^{2}}{2\sigma_{\mathrm{host}}^{2}}\right]},
\end{equation}
where $\mu$ and $\sigma_{\mathrm{host}}$ are the mean and standard deviation of $\ln{\mathrm{DM_{host}}}$, respectively. Both parameters are treated as free in the analysis.

Therefore, the total probability density function of a FRB can be determined as 

\begin{equation}
    P(\mathrm{DM_{ext}}(z_{i})) = \int_{0}^{\mathrm{DM_{ext}}} \mathrm{P_{IGM}}(\mathrm{DM_{IGM}}) \times \mathrm{P_{host}}(\mathrm{DM_{host}})  \, d\mathrm{DM_{host}},
\end{equation}
where $\mathrm{P_{IGM}}$ and $\mathrm{P_{host}}$ are given by Equations \ref{eq:pdf_1}, \ref{eq:pdf_2} and \ref{eq:pdf_3}, respectively. Finally, the joint likelihood function for all FRBs can be calculated by the relation

\begin{equation}
    \mathcal{L}_{\mathrm{tot}} = \prod_{i=1}^{N} P_{i}(\mathrm{DM_{ext,i}}(z_{i})),
\end{equation}
where the subscript $i$ represents the $i$-th FRBs data and $N$ is the
total number of FRBs data.

\begin{table}[htbp]
\centering
\caption{Summary of priors.}
\label{tab:priors}
\begin{tabular}{|c|c|c|}
\hline
\hline
Parameter & Prior & Range \\
\hline
$\Omega_{b}$ & Uniform &  $[0.0430, 0.0458]$ \\
$\Omega_{b}$  & Gaussian &  $\mathcal{N}(0.0444, 0.00028)$ \\
$H_{0}$ [km s$^{-1}$ Mpc$^{-1}$] & Uniform &  $[60, 80]$ \\
$q_{0}$ & Uniform &  $[-2, 1]$ \\
$j_0$ & Uniform &  $[-1, 2]$ \\
$f_{\mathrm{IGM,0}}$ & Uniform & $[0, 1]$\\
$\mu_{host}$ [pc/cm$^{3}$] & Uniform &  $[\ln(10), \ln(120)]$ \\
$\sigma_{host}$ & Uniform & $[0.001, 2.0]$ \\ 
\hline
\hline
\end{tabular}
\end{table}

As previously discussed, the baryon fraction in the IGM is still poorly constrained by observations and simulations, and it may be correlated with $\Omega_{b}$ and $H_{0}$ through Eq.~\ref{eq:igm}. To investigate the impact of this uncertainty on the cosmographic constraints, we consider two different treatments of $f_{\mathrm{IGM,0}}$. In the first approach, we fix its value to $f_{\mathrm{IGM,0}}=0.83$ \cite{Fukugita1998} (hereafter referred to as the fixed case). In the second approach, we treat $f_{\mathrm{IGM,0}}$ as a free parameter to be constrained by the data (hereafter referred to as the free case). We perform a Markov Chain Monte Carlo (MCMC) analysis using the \textit{MultiNest} algorithm \cite{MultiNest1,MultiNest2,MultiNest3} to constrain the model parameters. In the fixed case, the free parameters are $H_{0}$, $q_{0}$, $j_{0}$, $\mu_{\mathrm{host}}$, and $\sigma_{\mathrm{host}}$. In the free case, the parameter set is extended to include $f_{\mathrm{IGM,0}}$, resulting in the free parameters $H_{0}$, $q_{0}$, $j_{0}$, $f_{\mathrm{IGM,0}}$, $\mu_{\mathrm{host}}$, and $\sigma_{\mathrm{host}}$. Since, $\Omega_{b}$ has a parameter degeneracy with others parameters, we investigate the influence of this parameter through two scenario: first we consider a Uniform prior within the interval $[0.0430, 0.0458]$, which aligns with works reported in \cite{Planck2018,Cooke2018}; second, we adopt a Gaussian prior $\Omega_{b} = 0.0444 \pm 0.00028$, which is in agreement with \cite{Planck2018,Cooke2018}. In Table~\ref{tab:priors}, we report the prior adopted for each parameter and its corresponding range. We adopt wide prior intervals for the cosmographic parameters, especially for $q_{0}$ and $j_{0}$, enabling a robust exploration of the cosmological parameter space without imposing any assumptions about a specific underlying cosmological model. By doing that the chosen ranges are sufficiently broad to encompass most of the physically plausible parameter space, avoiding any a priori bias toward a particular cosmological scenario and ensuring that the resulting constraints are predominantly driven by the data.


\begin{figure*}
\begin{center}
\includegraphics[width=0.49\textwidth]{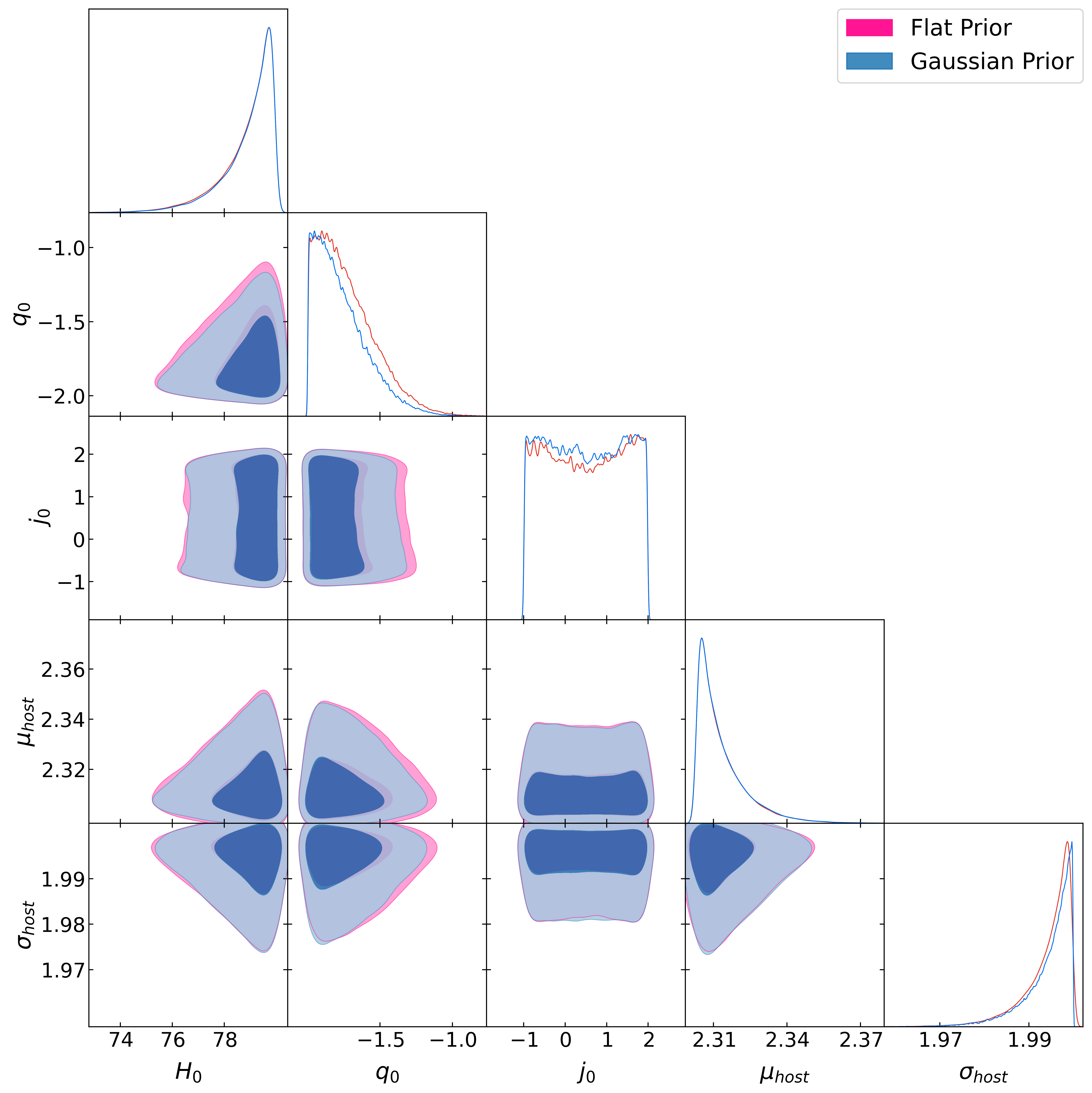}
\includegraphics[width=0.49\textwidth]{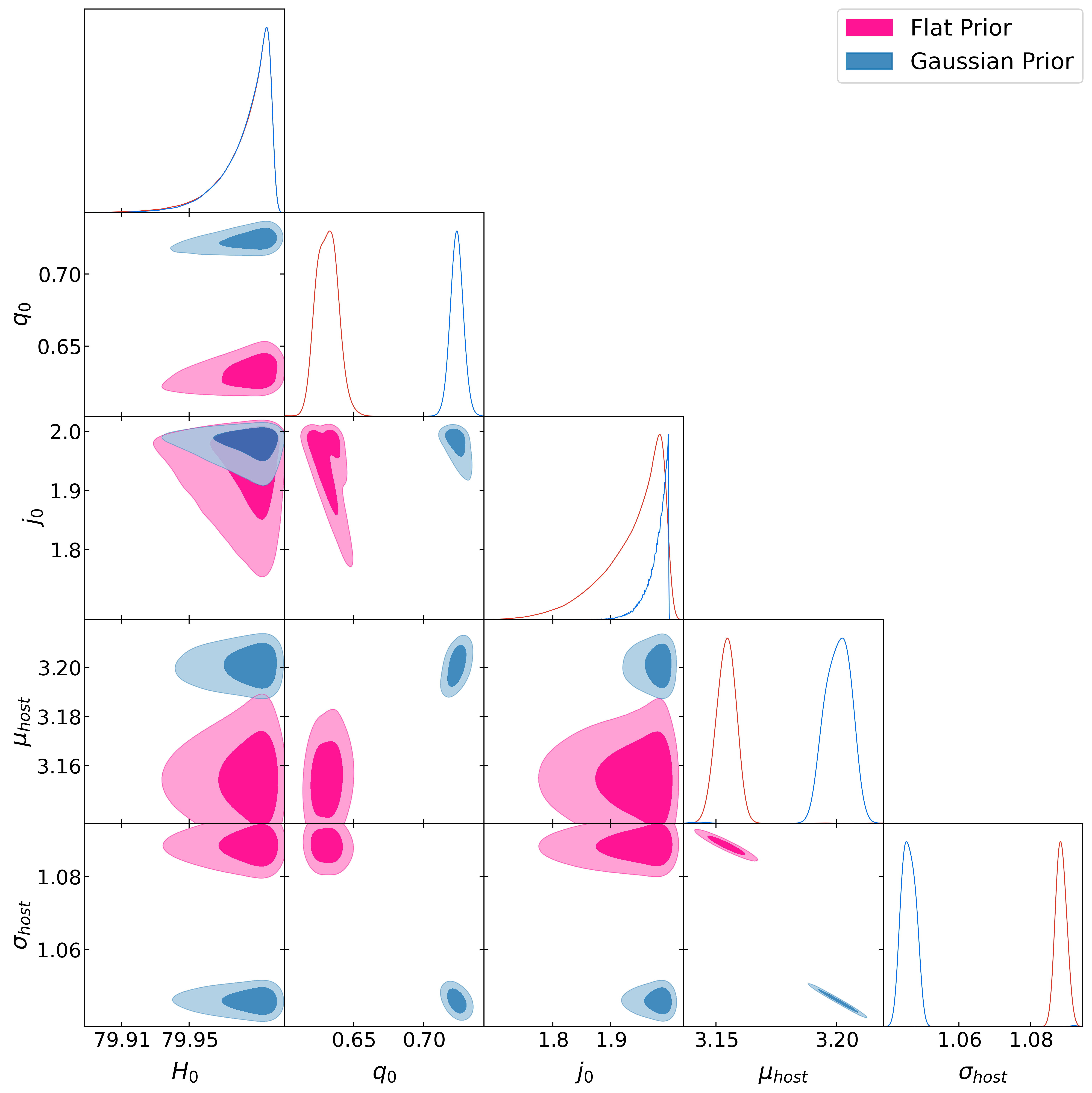}
\vspace{-0.5cm}
\caption{Constraints on the cosmographic parameters, as well as on the mean and variance of the host-galaxy contribution PDF, obtained for Distribution I (left panel) and Distribution II (right panel), considering a fixed $f_{\mathrm{IGM,0}}$. The pink and blue contours represent the $68\%$ and $95\%$ confidence intervals for the Flat and Gaussian priors on $\Omega_{b}$, respectively.}
\label{fig:f0fixed}
\end{center} 
\end{figure*}

\begin{figure*}
\begin{center}
\includegraphics[width=0.48\textwidth]{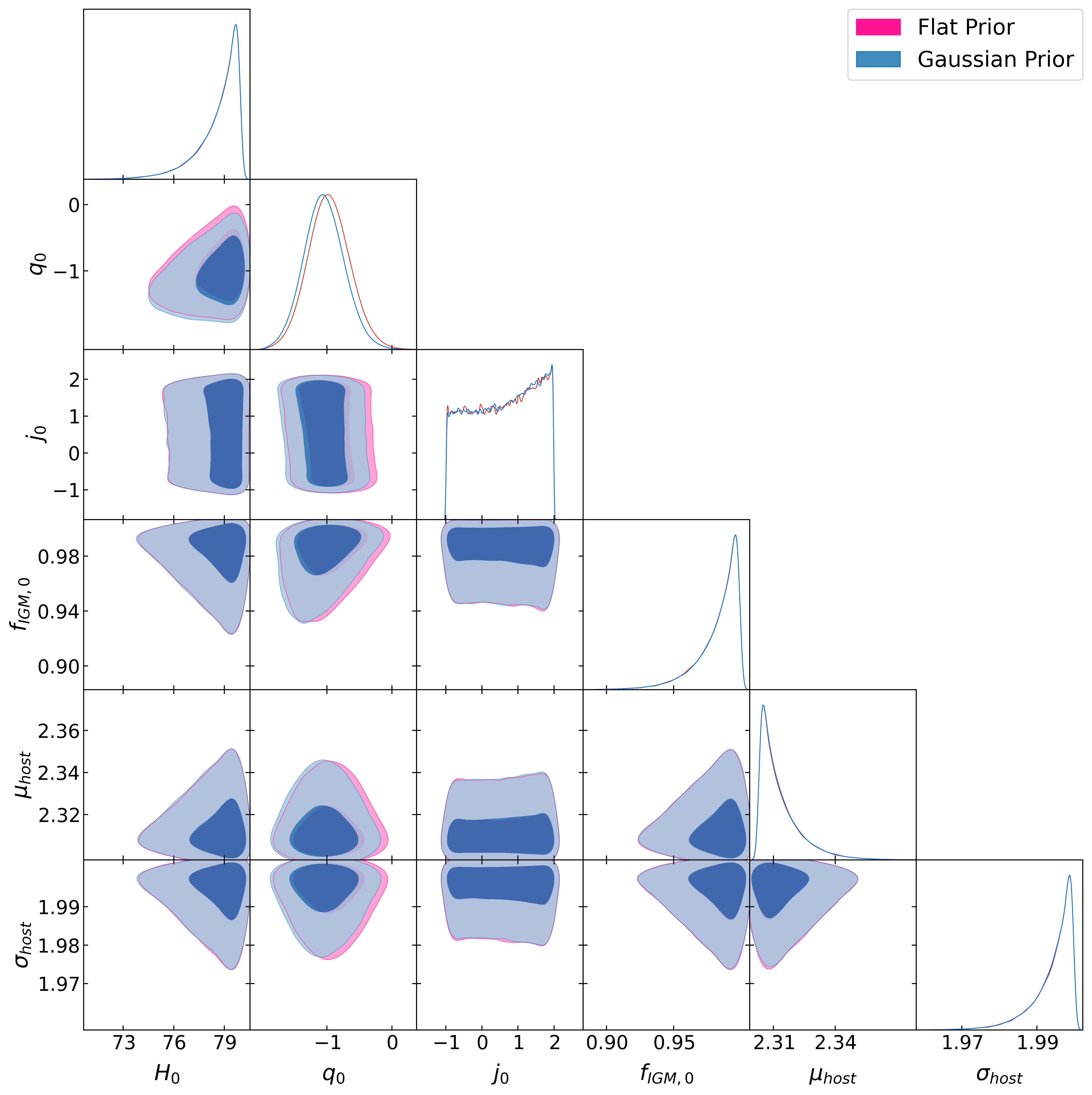}
\includegraphics[width=0.48\textwidth]{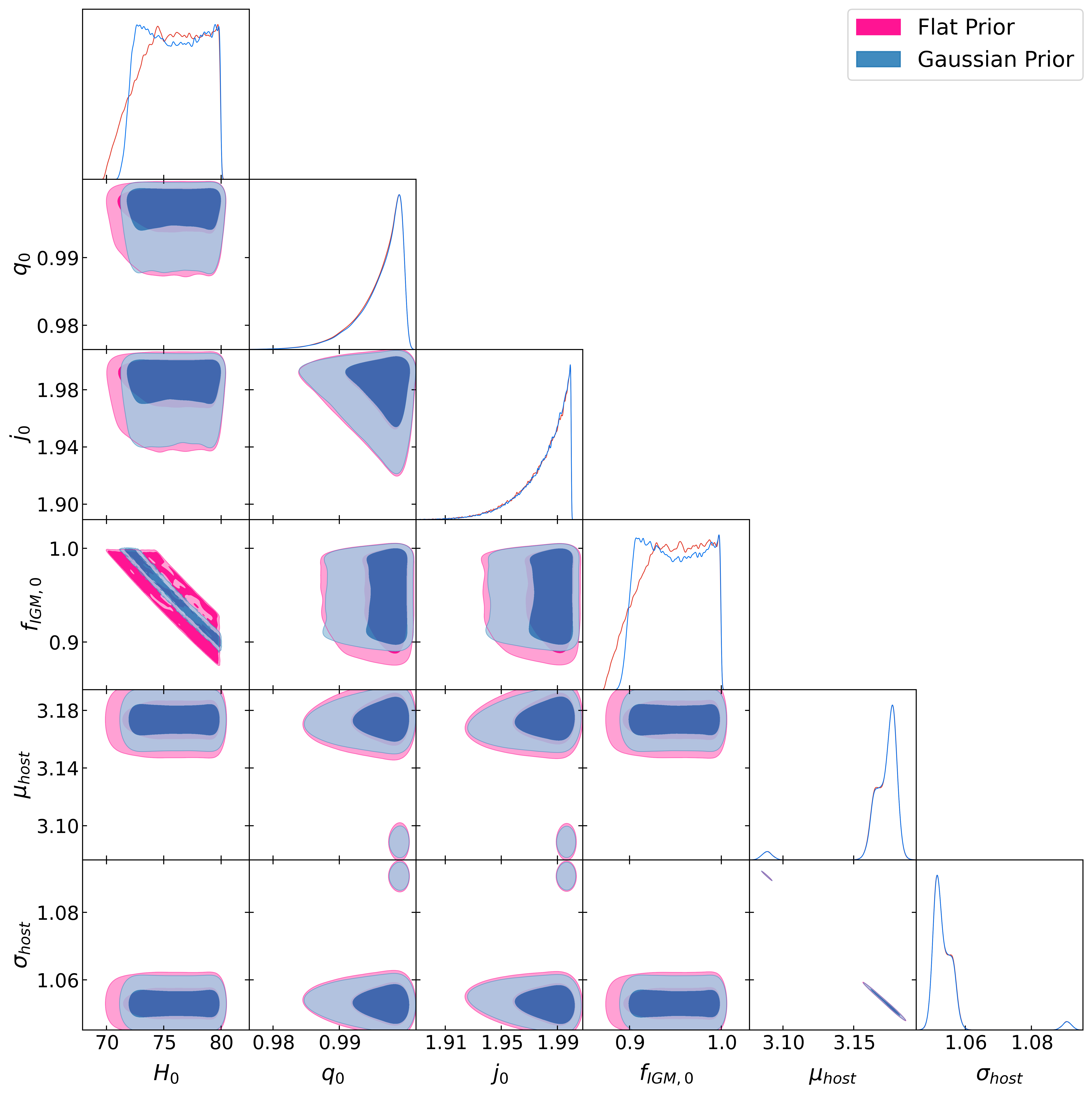}
\vspace{-0.5cm}
\caption{The same as in Fig.~\ref{fig:f0fixed}, but with $f_{\mathrm{IGM,0}}$ treated as a free parameter.}
\label{fig:f0free}
\end{center} 
\end{figure*}

\section{Results}
\label{sec:results}

\subsection{Case: $f_{\mathrm{IGM,0}}$ fixed}

\begin{table*}[h]
\vspace{0.3cm}
\centering
\caption{Estimates of the cosmographic parameters for fixed $f_{\mathrm{IGM,0}}$ (at $1 \sigma$).}
\begin{tabular}{ c  c  c  c c c  }
\hline
PDF & $H_{0}$ & $q_{0}$ & $j_{0}$ &  $\mu_{\mathrm{host}}$ & $\sigma_{\mathrm{host}}$ \\
 & [km/s/Mpc] &  &    &  [pc/cm$^{3}$] &  \\
\hline
\multicolumn{6}{c}{Uniform prior} \\
\hline
Distribution I   & $\geq 77.89$ & $-1.91^{+0.39}_{-0.01}$ & - & $\leq 2.32$& $\geq 1.99$  \\
Distribution II   & $\geq$ 79.97 & $0.633^{+0.006}_{-0.010}$ & $\geq 1.88$ & $3.155_{-0.005}^{+0.004}$ & $1.088\pm0.002$ \\
\hline
\hline
\multicolumn{6}{c}{Gaussian prior}  \\
\hline
Distribution I   & $\geq 78.02$ & $\leq -1.58$ & - & $\leq 2.32$& $\geq 1.99$  \\
Distribution II   & $\geq$ 79.97 & $0.724^{+0.004}_{-0.005}$ & $\geq 1.96$ & $3.202_{-0.007}^{+0.004}$ & $1.045_{-0.002}^{+0.003}$  \\
\hline
\end{tabular}
\label{tab:results_f0fixed}
\end{table*}

Figure \ref{fig:f0fixed} shows the posterior probability density functions and the $1$–$2\sigma$ contours for the combinations of the free parameters $H_{0}$, $q_{0}$, $j_{0}$, $\mu_{\mathrm{host}}$, and $\sigma_{\mathrm{host}}$, considering Distribution I from~\ref{eq:pdf_1} (left panel), and Distribution II from~\ref{eq:pdf_2} (right panel), with $f_{\mathrm{IGM,0}}$ fixed. Table \ref{tab:results_f0fixed} presents the constraints obtained for both priors and all distribution models. For parameters constrained only by one-sided limits, the inferred bounds are determined by the adopted prior range. Consequently, when the posterior reaches the edge of the prior, we report only the corresponding lower or upper limit rather than a two-sided confidence interval. For Distribution I, we estimate $q_{0} = -1.91^{+0.39}_{-0.01}$ for the flat prior and $q_{0} \leq -1.58$ for the Gaussian prior. The constraints on $j_{0}$ remain weak in all cases. The preference for positive $q_{0}$ in Distribution II may be attributed to a strong degeneracy between the cosmic expansion parameters and the host galaxy contribution, since marginalizing over $\mathrm{DM_{host,0}}$ may compensate for the intrinsic dispersion at low redshifts, shifting the deceleration parameter to reproduce the total observed DM budget.

Comparing our results for Distribution II with those obtained by Refs.~\cite{Fortunato2023,Lazaro}, which adopted the same IGM PDF, we find that the constraints on $q_{0}$ and $j_{0}$ are not compatible, as both works reported negative values of $q_{0}$ and tighter bounds on $j_{0}$. This discrepancy may be associated with the adopted prior ranges: in our analysis, we adopt broad priors to enable a more robust exploration of the parameter space without assuming a specific cosmological model, whereas those studies restricted the priors around the predictions of the $\Lambda$CDM model ($q_{0} = -0.55$ and $j_{0} = 1$). In the case of Ref.~\cite{Gao2024}, $q_{0}$ could not be constrained, preventing a direct comparison.

\subsection{Case: $f_{\mathrm{IGM,0}}$ free}

\begin{table*}[h]
\vspace{0.3cm}
\centering
\caption{Estimates of the cosmograpgic parametrs for free $f_{\mathrm{IGM,0}}$ (at $1 \sigma$).}
\begin{tabular}{ c  c  c  c c c c  }
\hline
PDF & $H_{0}$ & $q_{0}$ & $j_{0}$ & $f_{\mathrm{IGM,0}}$ & $\mu_{\mathrm{host}}$ & $\sigma_{\mathrm{host}}$ \\
 & [km/s/Mpc] &  &  &  &  [pc/cm$^{3}$] &  \\
\hline
\multicolumn{7}{c}{Uniform prior}  \\
\hline
Distribution I   & $\geq 77.49$& $-0.99^{+0.35}_{-0.30}$ & - & $\geq 0.97$ & $\leq 2.32$ & $\geq 1.99$ \\
Distribution II   & - & $\geq 0.993$ & $\geq 1.99$ & - & $3.176^{+0.006}_{-0.012}$ & $1.052^{+0.005}_{-0.002}$ \\
\hline
\hline
\multicolumn{7}{c}{Gaussian prior}  \\
\hline
Distribution I   & $\geq 77.47$& $-1.07^{+0.33}_{-0.29}$ & - & $\geq 0.97$ & $\leq 2.32$ & $\geq 1.99$ \\
Distribution II   & - & $\geq 0.993$ & $\geq 1.99$ & - & $3.176^{+0.006}_{-0.012}$ & $1.052^{+0.004}_{-0.002}$ \\
\hline
\end{tabular}
\label{tab:results_f0free}
\end{table*}

Figure \ref{tab:results_f0free}  shows the parametric space from the posterior probability density function and $1 - 2 \sigma$ contours for combinations of the parameters $H_{0}$, s$q_{0}$, $j_{0}$, $f_{\mathrm{IGM,0}}$, $\mu_{\mathrm{host}}$ and $\sigma_{\mathrm{host}}$. In Table \ref{tab:results_f0free} we present the corresponding best-fit results for the free-parameter case. As in the fixed case, for parameters exhibiting one-sided limits, the inferred bounds are driven by the adopted prior range. Therefore, we report only the corresponding lower or upper limit. For Distribution I, we estimate $q_{0} = -0.99^{+0.35}_{-0.30}$ (flat prior) and $q_{0} =-1.07^{+0.33}_{-0.29}$ (gaussian prior), while no meaningful constraints can be placed on $j_{0}$. Note that the inferred deceleration parameter here is higher than that obtained in the case where $f_{\mathrm{IGM,0}}$ is fixed (Tab.~\ref{tab:results_f0fixed}) and these values are in agreement with the  $\Lambda$CDM predictions at $2\sigma$ level. For the Distribution II, the best fits for both priors are the same, in contrast to the fixed case, although the constraints on $q_{0}$ remain positive. Overall, allowing $f_{\mathrm{IGM,0}}$ to vary improves the constraints for Distribution I, particularly on $q_{0}$, whereas for Distribution II, the deceleration parameter remains positive, but the results obtained with flat and Gaussian priors become more consistent with each other.

\section{New Prior}
\label{sec:newprior}

\begin{table}[htbp]
\centering
\caption{New priors.}
\label{tab:newprior}
\begin{tabular}{|c|c|c|}
\hline
\hline
Parameter & Prior & Range \\
\hline
$\Omega_{b}$ & Uniform &  $[0.0430, 0.0458]$ \\
$\Omega_{b}$  & Gaussian &  $\mathcal{N}(0.0444, 0.00028)$ \\
$H_{0}$ [km s$^{-1}$ Mpc$^{-1}$] & Uniform &  $[65, 75]$ \\
$q_{0}$ & Uniform &  $[-1, 0]$ \\
$j_0$ & Uniform &  $[0, 2]$ \\
$\mu_{host}$ [pc/cm$^{3}$] & Uniform &  $[\ln(10), \ln(120)]$ \\
$\sigma_{host}$ & Uniform & $[0.001, 2.0]$ \\ 
\hline
\hline
\end{tabular}
\end{table}

\begin{table*}[h]
\vspace{0.3cm}
\centering
\caption{Same as Tab.~\ref{tab:results_f0fixed}, but adopting the prior ranges listed in Tab.~\ref{tab:newprior} (at $1 \sigma$).}
\begin{tabular}{ c  c  c  c c c   }
\hline
PDF & $H_{0}$ & $q_{0}$ & $j_{0}$  & $\mu_{\mathrm{host}}$ & $\sigma_{\mathrm{host}}$ \\
 & [km/s/Mpc] &  &  &    [pc/cm$^{3}$] &  \\
\hline
\multicolumn{6}{c}{Uniform prior}  \\
\hline
Distribution I   & $\geq 73.87$& $\leq-0.84$ & - & $\leq 2.5$ & $\geq 1.89$ \\
Distribution II   & - & $-0.526^{+0.027}_{-0.003}$ & $\geq 1.13$ & $3.430^{+0.003}_{-0.001}$ & $\geq 1.00$ \\
\hline
\hline
\multicolumn{6}{c}{Gaussian prior}  \\
\hline
Distribution I   & $\geq 73.91$& $\leq-0.85$ & - & $\leq 2.1$ & $\geq 1.89$ \\
Distribution II  & $66.81^{+0.45}_{-0.36}$ & $-0.525^{+0.026}_{-0.003}$ & $\geq 1.15$ & $3.430^{+0.003}_{-0.001}$ & $\geq 1.00$ \\
\hline
\end{tabular}
\label{tab:results_newprior}
\end{table*}

\begin{figure*}
\begin{center}
\includegraphics[width=0.49\textwidth]{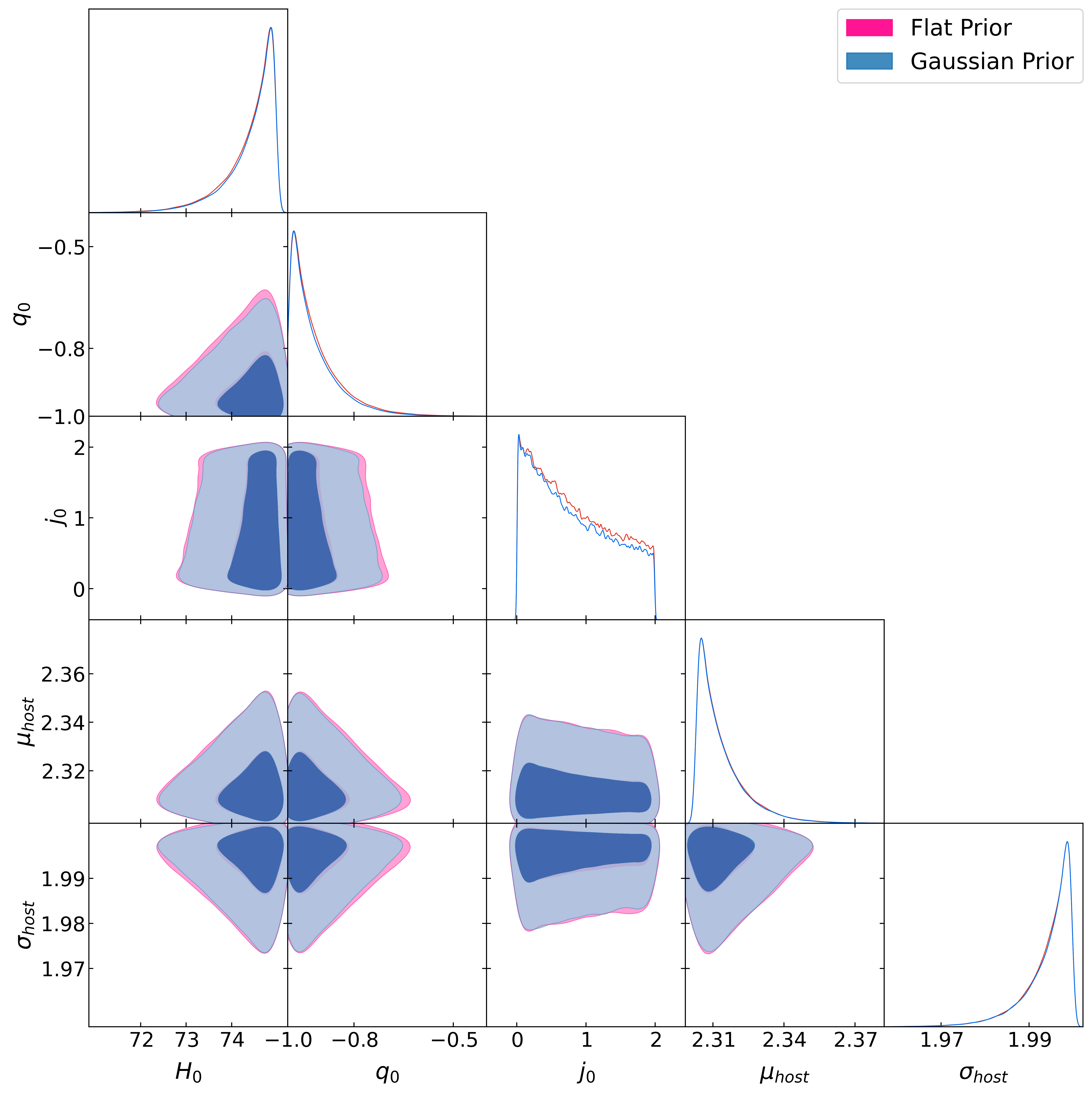}
\includegraphics[width=0.49\textwidth]{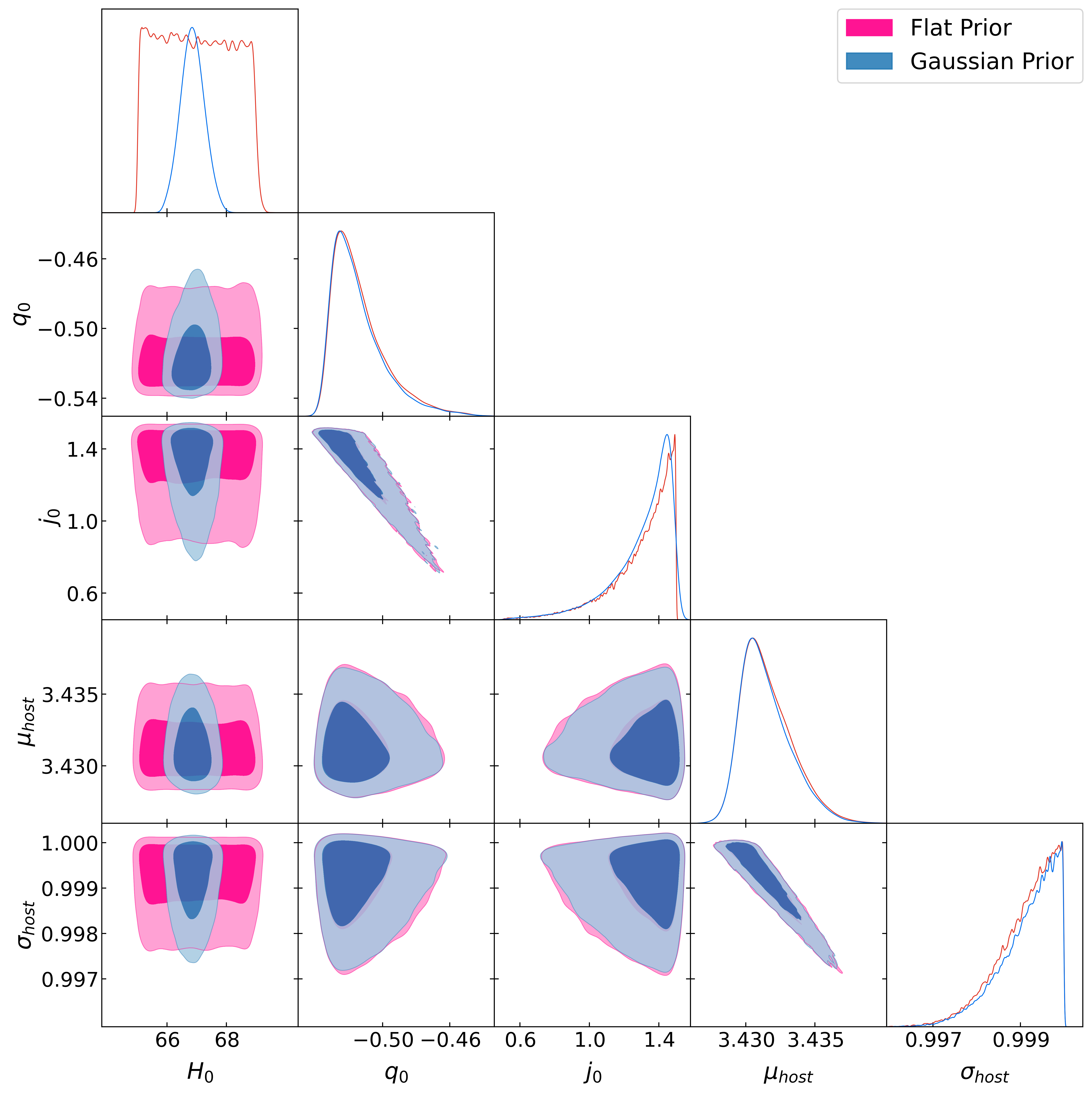}
\vspace{-0.5cm}
\caption{The same as in Fig.~\ref{fig:f0fixed} considering the prior ranges listed in Tab.~\ref{tab:newprior}.}
\label{fig:newprior}
\end{center} 
\end{figure*}

In the previous section, we presented the constraints on the cosmographic parameters adopting wide prior ranges, particularly for $q_{0}$ and $j_{0}$, and we found that, for Distribution II, the deceleration parameter assumes positive values, in contrast with the negative results reported by Refs.~\cite{Fortunato2023,Lazaro}, suggesting that the choice of priors on $q_{0}$ may significantly influence the outcome. To evaluate the impact of these prior assumptions on the constraints presented in Section~\ref{sec:results}, we redo the analysis for the fixed $f_{\mathrm{IGM,0}}$ case, modifying the prior ranges. Whereas our baseline analysis adopted wide priors to enable a robust and model-independent exploration of the parameter space, we now impose more restrictive priors, similar to those used in previous studies, in order to assess how prior choices affect the inferred cosmographic constraints. The new prior ranges are summarized in Table~\ref{tab:newprior}.

The results of our analysis are displayed in Fig. \ref{fig:newprior} and Table \ref{tab:results_newprior}. Figure \ref{fig:newprior} shows the posterior probability density function and $1-2\sigma$ contours on the parametric spaces for the new prior range. The quantitative results of the analysis, displayed in Table \ref{tab:results_newprior}, show the impact of the priors in the determination of the cosmographic parameters, particularly  $q_{0}$. While for Distribution I, we obtain $q_{0} \lesssim -0.84$, which is higher than the obtained for the original prior (Tab. \ref{tab:results_f0fixed}), for Distribution II we estimate $q_{0}$ negative, $q_{0} = -0.526^{+0.027}_{-0.003}$ (Flat prior) and $q_{0} = -0.525^{+0.026}_{-0.003}$ (Guassian prior), being both values in agrrement with the results reported by  Refs.~\cite{Fortunato2023,Lazaro} and in agreement with predictions of the $\Lambda$CDM model, $q_{0} = -0.55$. In the case of the jerk parameter for Distribution II,  we obtain $j_{0} \gtrsim 1.13$ for both prior settings. 

It is worth emphasizing that, even when adopting tighter priors and considering the baryon fraction in IGM as a free parameter, the jerk parameter cannot be strongly constrained with the current FRB sample, differently of previous works \cite{Fortunato2023,Lazaro}. This behavior is expected, given the significant scatter of the data points and their relatively large error bars (see Fig.~\ref{fig:data}), which limit the sensitivity of the analysis to higher-order cosmographic terms.


\section{Conclusion and discussion}
\label{sec:conclusion}

Despite the remarkable success of the $\Lambda$CDM model, the nature of the mechanism driving the late-time acceleration of the Universe remains uncertain. In this context, cosmography offers a robust framework for inferring kinematic parameters, such as the Hubble constant, the deceleration parameter, and the jerk parameter, without relying on a specific cosmological model.

In this paper, we apply the cosmographic framework to constrain these kinematic parameters using Fast Radio Bursts (FRBs). The original sample comprises 117 FRBs; after applying quality-selection criteria, the final dataset consists of 106 events spanning the redshift range $z \leq 0.7$. Since cosmographic expansions in terms of the redshift $z$ may exhibit convergence issues at moderate redshifts, we adopt the $y$-redshift parametrization, $y=z/(1+z)$, and truncate the cosmographic series at third order.

Since FRBs are not completely understood, two other aspects are important to be addressed: The host galaxy and the intergalactic medium. By following the literature, we assume a log-normal distribution for the host galaxy, and for the IGM, we assume two different distributions: a Gaussian distribution with a fixed fluctuation (Distribution I), and a quasi-Gaussian distribution with parameters that map the inhomogeneity of the intergalactic medium. Furthermore, we analyse the impact of different assumptions on the baryon mass fraction in the IGM, specifically one with $f_{\mathrm{IGM},0}$ fixed at its standard observational value and another considering it as a free parameter.

The results when $f_{\mathrm{IGM},0}$ is fixed for Distribution I yield an upper limit for the Hubble constant around $H_0 \sim 78 \; km/s/Mpc$ when we assume both uniform and Gaussian prior for the baryonic density parameter. The jerk parameter remains weak in both cases, but the deceleration parameter shows different results, numerically with $q_{0} = -1.91^{+0.39}_{-0.01}$ for the flat prior and $q_{0} \leq -1.58$ for the Gaussian prior. While for Distribution II, there is a preference for positive $q_0$, probably due to a degeneracy between the cosmic expansion parameters and the host galaxy contribution.

On the other hand, when $f_{\mathrm{IGM},0}$ is free, the results for the Hubble constant and the jerk parameter do not change significantly, but the deceleration parameter is slightly different. When we assume the Distribution I we obtain $q_{0} = - 0.99_{- 0.30}^{+0.35}$ (for a flat prior) and $q_{0} = - 1.07_{- 0.29}^{+0.33}$ (for a gaussian prior).

Alternatively, we also explore a different set of priors with a more restrictive range of values, as shown in Table~\ref{tab:newprior}. The main result of these analysis is in the determination of the deceleration parameter, since we obtain $q_{0} \lesssim -0.84$ for Distribution I, and $q_{0} = -0.526^{+0.027}_{-0.003}$ (Flat prior) and $q_{0} = -0.525^{+0.026}_{-0.003}$ (Guassian prior), for Distribution II.

Our analyses show a clear dependence of the cosmographic parameters when we explore the Fast Radio Burst in the context of a cosmological model-independent approach. We show that the probability distribution of the IGM dispersion measure is a key factor as well as the choice of prior, for both cosmological and kinematic parameters. Since we expect a growing number of FRB detections, the methodology used in this paper can be used as a motivation and complement to multi-messenger observations showing the full potential of FRBs as cosmological probes of high precision.


\section*{Acknowledgements}

TL thanks the financial support from the Conselho Nacional de Desenvolvimento Cient\'{\i}fico e Tecnol\'ogico (CNPq). RSG thanks financial support from the Funda\c{c}\~ao de Amparo \`a Pesquisa do Estado do Rio de Janeiro (FAPERJ) grant SEI-260003/005977/2024 - APQ1. JSA is supported by Conselho Nacional de Desenvolvimento Cient\'{\i}fico e Tecnol\'ogico (CNPq No. 307683/2022-2; CNPq No. 448158/2025-6) and Funda\c{c}\~ao de Amparo \`a Pesquisa do Estado do Rio de Janeiro (FAPERJ) grant 259610 (2021). This work was developed thanks to the High-Performance Computing Center at the National Observatory (CPDON).




\newpage
\begin{longtable}[c]{l c c c c l}
\caption{Properties of well-localized FRBs}\\
\hline
\hline
\hline
Name & $z$ & $\mathrm{DM}_{\mathrm{ISM}}$ & $\mathrm{DM}_{\mathrm{obs}}$ & $\sigma_{\mathrm{obs}}$ & Refs. \\
& & [pc/cm$^{3}$] & [pc/cm$^{3}$] & [pc/cm$^{3}$] & \\
\hline
\endfirsthead

\hline
\hline
\hline
Name & $z$ & $\mathrm{DM}_{\mathrm{ISM}}$ & $\mathrm{DM}_{\mathrm{obs}}$ & $\sigma_{\mathrm{obs}}$ & Refs. \\
& & [pc/cm$^{3}$] & [pc/cm$^{3}$] & [pc/cm$^{3}$] & \\
\hline
\endhead

\hline
\endfoot

\hline
FRB 20121102A	&	0.19273	&	188.0	&	557.0	&	2.0	&	\cite{FRB121102}	\\
FRB 20180301A	&	0.3305	&	152.0	&	536.0	&	8.0	&	\cite{FRB20191228}	\\
FRB 20180814	&	0.068	&	87.75	&	189.4	&	0.4	&	\cite{FRB180814}	\\
FRB 20180916B	&	0.0337 	&	200.0	&	348.80	&	0.2	&	\cite{FRB180916}	\\
FRB 20180924B	&	0.3214	&	40.5	&	361.42	&	0.06	&	\cite{FRB180924}	\\
FRB 20181112A	&	0.4755	&	102.0	&	589.27	&	0.03	&	\cite{FRB181112}	\\
FRB 20181220A	&	0.2746	&	122.81	&	208.66	&	1.62	&	\cite{FRB181220A}	\\
FRB 20181223C	&	0.03024	&	19.9	&	112.45	&	0.01	&	\cite{FRB181220A}	\\
FRB 20190102C	&	0.2913	&	57.3	&	363.6	&	0.3	&	\cite{FRB190102}	\\
FRB 20190110C	&	0.12244	&	37	&	221.92	&	0.01	&	\cite{FRB190110C}	\\
FRB 20190303A	&	0.064	&	29.39	&	222.4	&	0.7	&	\cite{FRB180814}	\\
FRB 20190418A	&	0.07132	&	70.2	&	182.78	&	1.62	&	\cite{FRB181220A}	\\
FRB 20190425A	&	0.03122	&	49.25	&	127.78	&	1.62	&	\cite{FRB181220A}	\\
FRB 20190523A	&	0.66	&	37.0	&	760.8	&	0.6	&	\cite{FRB190523_1,FRB190523_2}	\\
FRB 20190608B	&	0.1178	&	37.2	&	338.7	&	0.5	&	\cite{FRB190608}	\\
FRB 20190611B	&	0.378	&	57.83	&	321.4	&	0.2	&	\cite{FRB190523_2}	\\
FRB 20190711A	&	0.522	&	56.4	&	593.1	&	0.4	&	\cite{FRB190523_2}	\\
FRB 20190714A	&	0.2365	&	38.0	&	504.13	&	2.0	&	\cite{FRB190523_2}	\\
FRB 20191001A	&	0.234	&	44.7	&	506.92	&	0.04	&	\cite{FRB190523_2}	\\
FRB 20191106C	&	0.10775	&	25	&	333.40	&	0.2	&	\cite{FRB190110C}	\\
FRB 20191228A	&	0.2432	&	33.0	&	297.5	&	0.05	&	\cite{FRB20191228}	\\
FRB 20200223B	&	0.06024	&	46	&	202.268	&	0.007	&	\cite{FRB190110C}	\\
FRB 20200430A	&	0.16	&	27.0	&	380.25	&	0.5	&	\cite{FRB190523_2}	\\
FRB 20200906A	&	0.3688	&	36	&	577.8	&	0.02	&	\cite{FRB20191228}	\\
FRB 20201123A	&	0.0507	&	251.93	&	433.55	&	0.0036	&	\cite{FRB201123A}	\\
FRB 20201124A	&	0.098	&	123.2	&	413.52	&	0.5	&	\cite{FRB201124}	\\
FRB 20210117A	&	0.2145	&	34.4	&	730.0	&	1.0	&	\cite{FRB210117A}	\\
FRB 20210320	&	0.2797	&	42.2	&	384.8	&	0.3	&	\cite{FRB210320}	\\
FRB 20210410D	&	0.1415	&	56.2	&	578.78	&	2.0	&	\cite{FRB210410D}	\\
FRB 20210603A 	&	0.1772	&	40.0	&	500.147	&	0.004	&	\cite{FRB210603A}	\\
FRB 20210807D	&	0.12927	&	121.2	&	251.9	&	0.2	&	\cite{FRB210807D}	\\
FRB 20211127I	&	0.0469	&	42.5	&	234.83	&	0.08	&	\cite{FRB210807D}	\\
FRB 20211203C 	&	0.3439	&	63.4	&	636.2	&	0.4	&	\cite{FRB211203C}	\\
FRB 20211212A	&	0.0715	&	27.1	&	206.0	&	5.0	&	\cite{FRB210807D}	\\
FRB 20220105A	&	0.2785	&	22.0	&	583	&	1.0	&	\cite{FRB211203C}	\\
FRB 20220204A	&	0.4	&	50.7	&	612.2	&	0.05	&	\cite{FRB220204A}	\\
FRB 20220207C	&	0.043040	&	79.3 	&	262.38	&	0.01	&	\cite{FRB220207C}	\\
FRB 20220208A	&	0.351	&	101.6	&	437	&	0.6	&	\cite{FRB220204A}	\\
FRB 20220307B	&	0.248123	&	135.7 	&	499.27 	&	0.06	&	\cite{FRB220207C}	\\
FRB 20220310F	&	0.477958	&	45.4 	&	462.24	&	0.005	&	\cite{FRB220207C}	\\
FRB 20220330D 	&	0.3714	&	38.6	&	468.1	&	0.85	&	\cite{FRB220204A}	\\
FRB 20220418A	&	0.622000	&	37.6 	&	623.25	&	0.01	&	\cite{FRB220207C}	\\
FRB 20220501C	&	0.381	&	31	&	449.5	&	0.2	&	\cite{FRB210807D}	\\
FRB 20220506D	&	0.30039 	&	89.1	&	396.97	&	0.02	&	\cite{FRB220207C}	\\
FRB 20220509G	&	0.089400	&	55.2	&	269.53	&	0.02	&	\cite{FRB220207C}	\\
FRB 20220529A	&	0.1839	&	40	&	246	&	$\dag$	&	\cite{FRB220529A}	\\
FRB 20220717A	&	0.36295	&	118	&	637.34	&	3.52	&	\cite{FRB220717A}	\\
FRB 20220725A	&	0.1926	&	31	&	290.4	&	0.3	&	\cite{FRB210807D}	\\
FRB 20220726A	&	0.361	&	89.5	&	686.55	&	0.01	&	\cite{FRB220204A}	\\
FRB 20220825A	&	0.241397	&	79.7	&	651.24	&	0.06	&	\cite{FRB220207C}	\\
FRB 20220831A 	&	0.262	&	1019.50	&	1146.25	&	0.2	&	\cite{FRB220204A}	\\
FRB 20220912A	&	0.0771	&	125.00	&	219.46	&	0.042	&	\cite{FRB220912A}	\\
FRB 20220914A	&	0.113900 	&	55.2	&	631.28	&	0.04	&	\cite{FRB220207C}	\\
FRB 20220918A	&	0.491	&	41	&	656.8	&	0.8	&	\cite{FRB210807D}	\\
FRB 20220920A	&	0.158239	&	40.3	&	314.99 	&	0.01	&	\cite{FRB220207C}	\\
FRB 20221012A	&	0.284669	&	54.4	&	441.08	&	0.7	&	\cite{FRB220207C}	\\
FRB 20221027A 	&	0.229	&	47.2	&	452.5	&	$\dag$	&	\cite{FRB220204A}	\\
FRB 20221101B	&	0.2395	&	131.2	&	490.7	&	$\dag$ &	\cite{FRB220204A}	\\
FRB 20221106A	&	0.2044	&	35	&	343.8	&	0.8	&	\cite{FRB210807D}	\\
FRB 20221113A	&	0.2505	&	91.7	&	411.4	&	$\dag$ &	\cite{FRB220204A}	\\
FRB 20221116A 	&	0.2764	&	132.3	&	640.6	&	$\dag$	&	\cite{FRB220204A}	\\
FRB 20221219A 	&	0.554	&	44.4	&	706.7	&	0.6	&	\cite{FRB221219A}	\\
FRB 20230124 	&	0.094	&	38.5	&	590.6	&	$\dag$ &	\cite{FRB220204A}	\\
FRB 20230203A	&	0.1464	&	36.29	&	420.1	&	$\dag$ &	\cite{FRB230203A}	\\
FRB 20230216A 	&	0.5310	&	38.5	&	828	&	$\dag$ &	\cite{FRB220204A}	\\
FRB 20230222A	&	0.1223	&	134.13	&	706.1	&	$\dag$	&	\cite{FRB230203A}	\\
FRB 20230222B	&	0.11	&	27.7	&	187.8	&	$\dag$	&	\cite{FRB230203A}	\\
FRB 20230307A 	&	0.2710	&	37.6	&	608.9	&	$\dag$	&	\cite{FRB220204A}	\\
FRB 20230311A	&	0.1918	&	92.39	&	364.3	&	$\dag$	&	\cite{FRB230203A}	\\
FRB 20230501A 	&	0.3010	&	125.6	&	532.50	&	$\dag$	&	\cite{FRB220204A}	\\
FRB 20230526A	&	0.1570	&	50	&	361.4	&	0.2	&	\cite{FRB210807D}	\\
FRB 20230626A 	&	0.327	&	39.2	&	451.2	&	$\dag$	&	\cite{FRB220204A}	\\
FRB 20230628A	&	0.1265	&	39.1	&	345.15	& $\dag$	&	\cite{FRB220204A}	\\
FRB 20230703A	&	0.1184	&	26.97	&	291.3	&	$\dag$	&	\cite{FRB230203A}	\\
FRB 20230708A	&	0.105	&	50	&	411.51	&	0.05	&	\cite{FRB210807D}	\\
FRB 20230712A	&	0.4525	&	39.2	&	586.96	&	$\dag$	&	\cite{FRB220204A}	\\
FRB 20230718A	&	0.035	&	396	&	477.0	&	0.5	&	\cite{FRB210807D}	\\
FRB 20230730A	&	0.2115	&	85.18	&	312.5	&	$\dag$	&	\cite{FRB230203A}	\\
FRB 20230814A	&	0.5535	&	104.9	&	696.35	&	0.5	&	\cite{FRB220204A}	\\
FRB 20230902A	&	0.3619	&	34	&	440.1	&	0.1	&	\cite{FRB210807D}	\\
FRB 20230926A	&	0.0553	&	52.69	&	222.8	&	$\dag$	&	\cite{FRB230203A}	\\
FRB 20231005A	&	0.0713	&	33.37	&	189.4	&	$\dag$	&	\cite{FRB230203A}	\\
FRB 20231011A	&	0.0783	&	70.36	&	186.3	&	$\dag$	 &	\cite{FRB230203A}	\\
FRB 20231017A 	&	0.245	&	64.55	&	344.2	&	$\dag$	&	\cite{FRB230203A}	\\
FRB 20231025B	&	0.3238	&	48.67	&	368.7	&	$\dag$	&	\cite{FRB230203A}	\\
FRB 20231120A 	&	0.07	&	43.8	&	438.9	&	$\dag$	&	\cite{FRB220204A}	\\
FRB 20231123A	&	0.0729	&	89.76	&	302.1	&	$\dag$	&	\cite{FRB230203A}	\\
FRB 20231123B 	&	0.2625	&	40.2	&	396.7	&	$\dag$	&	\cite{FRB220204A}	\\
FRB 20231128A	&	0.1079	&	25.05	&	331.6	&	$\dag$	&	\cite{FRB230203A}	\\
FRB 20231201A	&	0.1119	&	70.03	&	169.4	&	$\dag$	&	\cite{FRB230203A}	\\
FRB 20231204A	&	0.0644	&	29.73	&	221.0	&	$\dag$	&	\cite{FRB230203A}	\\
FRB 20231206A	&	0.0659	&	59.13	&	457.7	&	$\dag$	&	\cite{FRB230203A}	\\
FRB 20231220A	&	0.3355	&	49.9	&	491.20	&	$\dag$	&	\cite{FRB220204A}	\\
FRB 20231223C	&	0.1059	&	47.9	&	165.8	&	$\dag$	&	\cite{FRB230203A}	\\
FRB 20231226A	&	0.1569	&	145	&	329.9	&	0.1	&	\cite{FRB210807D}	\\
FRB 20231229A	&	0.0190	&	58.12	&	198.5	&	$\dag$	&	\cite{FRB230203A}	\\
FRB 20231230A	&	0.0298	&	61.51	&	131.4	&	$\dag$	&	\cite{FRB230203A}	\\
FRB 20240114A	&	0.13	&	49.7	&	527.65	&	0.01	&	\cite{FRB240114A}	\\
FRB 20240119A	&	0.37	&	37.9	&	483.10	&	$\dag$	&	\cite{FRB220204A}	\\
FRB 20240201A	&	0.042729	&	38	&	374.5	&	0.2	&	\cite{FRB210807D}	\\
FRB 20240209A	&	0.1384	&	55.5	&	176.49	&	0.01	&	\cite{FRB240209A}	\\
FRB 20240210A	&	0.023686	&	31	&	283.73	&	0.05	&	\cite{FRB210807D}	\\
FRB 20240213A	&	0.1185	&	40.1	&	357.4	&	$\dag$	&	\cite{FRB220204A}	\\
FRB 20240215A 	&	0.21	&	48.0	&	549.5	&	$\dag$	&	\cite{FRB220204A}	\\
FRB 20240229A	&	0.2870	&	37.9	&	491.15	&	$\dag$	&	\cite{FRB220204A}	\\
FRB 20240310A	&	0.1270	&	36	&	601.8	&	0.2	&	\cite{FRB210807D}	\\
\label{tab:data}
\end{longtable}

\end{document}